\documentclass[journal=jacsat,manuscript=article]{achemso}

\usepackage[version=3]{mhchem} 

\usepackage{graphicx}%
\usepackage{multirow}%
\usepackage{amsmath,amssymb,amsfonts}%
\usepackage{amsthm}%
\usepackage{mathrsfs}%
\usepackage[title]{appendix}%
\usepackage{xcolor}%
\usepackage{textcomp}%
\usepackage{manyfoot}%
\usepackage{booktabs}%
\usepackage{algorithm}%
\usepackage{algorithmicx}%
\usepackage{algpseudocode}%
\usepackage{listings}%
\usepackage{bm}
\usepackage{physics}
\usepackage{tabularx}
\usepackage{hyperref}
\theoremstyle{thmstyleone}%
\theoremstyle{thmstyletwo}%

\theoremstyle{thmstylethree}%

\author{Wei Liu}
\affiliation{Department of Chemistry, School of Science and Research Center for Industries of the Future, Westlake University, Hangzhou, Zhejiang 310030, China}
\altaffiliation{Institute of Natural Sciences, Westlake Institute for Advanced Study, Hangzhou, Zhejiang 310024, China}
\author{Wenjie Dou}
\affiliation{Department of Chemistry, School of Science and Research Center for Industries of the Future, Westlake University, Hangzhou, Zhejiang 310030, China}
\altaffiliation{Institute of Natural Sciences, Westlake Institute for Advanced Study, Hangzhou, Zhejiang 310024, China}
\email{wenjiedou@westlake.edu.cn}

\title{Implicitly Restarted Lanczos Enables Chemically-Accurate Shallow Neural Quantum States}

\abbreviations{IR,NMR,UV}

\begin{document}

\begin{tocentry}
\includegraphics[height=3.6cm]{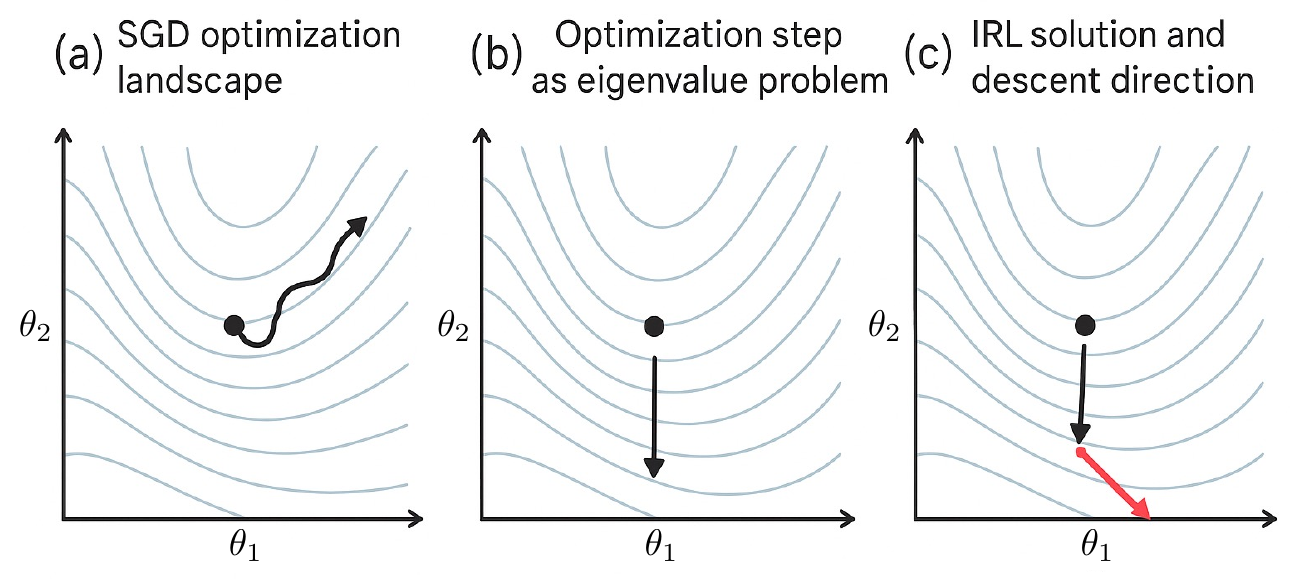}

\end{tocentry}

\begin{abstract}
  The variational optimization of high-dimensional neural network models, such as those used in neural quantum states (NQS), presents a significant challenge in machine intelligence. Conventional first-order stochastic methods (e.g., Adam) are plagued by slow convergence, sensitivity to hyperparameters, and numerical instability, preventing NQS from reaching the high accuracy required for fundamental science. We address this fundamental optimization bottleneck by introducing the implicitly restarted Lanczos (IRL) method as the core engine for NQS training. Our key innovation is an inherently stable second-order optimization framework that recasts the ill-conditioned parameter update problem into a small, well-posed Hermitian eigenvalue problem. By solving this problem efficiently and robustly with IRL, our approach automatically determines the optimal descent direction and step size, circumventing the need for demanding hyperparameter tuning and eliminating the numerical instabilities common in standard iterative solvers. We demonstrate that IRL enables shallow NQS architectures (with orders of magnitude fewer parameters) to consistently achieve extreme precision (1e-12 kcal/mol) in just 3 to 5 optimization steps. For the $\text{F}_2$ molecule, this translates to an approximate 17,900-fold speed-up in total runtime compared to Adam. This work establishes IRL as a superior, robust, and efficient second-order optimization strategy for variational quantum models, paving the way for the practical, high-fidelity application of neural networks in quantum physics and chemistry.
\end{abstract}

\section{Introduction}
The emergence of neural quantum states (NQS) has provided a powerful new, scalable paradigm for representing quantum many-body wave functions~\cite{carleo2017solving,lange2024architectures,gu2025solving,li2025spin,fischer2012introduction,netket2:2019,choo2020fermionic,schmitt2020quantum,hibat2021variational,vicentini2022netket,pederson2022machine,reh2023optimizing,passetti2023can,wu2024variational,liu2025absorption,bodendorfer2025variational,zhao2023scalable,wu2023nnqs,ren2023towards,han2020solving,brown1996combining}. Highly accurate predictions from NQS—which are intrinsically flexible trial wave functions optimized via variational~\cite{bodendorfer2025variational,liu2025absorption,wu2023nnqs,zhao2023scalable} or diffusion~\cite{ren2023towards,han2020solving,brown1996combining} Monte Carlo (VMC/DMC)—are contingent upon both the NQS architecture's sufficient expressive power and the effective optimization of its variational parameters. However, despite the architectural flexibility of NQS, achieving chemically accurate results faces significant optimization hurdles. Conventional first-order stochastic methods, such as stochastic gradient descent (SGD) and Adam~\cite{kingma2014adam}, are widely used but suffer from painfully slow convergence, high sensitivity to hyperparameters, and numerical instabilities in flat energy landscapes. Conversely, more powerful second-order algorithms—including the Newton \cite{umrigar2005energy}, approximate Newton \cite{sorella2005wave}, linear method (LM) \cite{nightingale2001optimization,toulouse2007optimization,toulouse2008full,umrigar2007alleviation}, and stochastic reconfiguration (SR) \cite{sorella2001generalized,shulenburger2008correlation,neuscamman2012optimizing} methods—are generally more efficient but suffer from a major limitation in the modern NQS era. These methods are currently constrained by their need to explicitly build and store large matrices, such as the quantum geometric tensor (QGT) $S$ and the effective Hamiltonian $H_{\text{eff}}$, whose size scales quadratically with the number of variational parameters ($O(P^2)$). Given that modern NQS typically employ tens of thousands to millions of parameters, this quadratic scaling renders these second-order approaches computationally intractable due to both immense memory consumption and prohibitive wall-clock time. Thus, to make NQS truly effective and circumvent the computational burden of massive architectures, it is imperative to develop optimization methods that achieve the efficiency of second-order techniques without suffering from their memory and scaling limitations. 

The LM and SR optimization methods, which are foundational second-order techniques in NQS, reduce the optimization task to solving either a system of linear equations or a linear eigenvalue problem. The matrices involved ($S$ and $H_{\text{eff}}$) are determined by stochastic sampling within the VMC framework. The essential difficulty, as discussed, is that the dimension of these matrices equals the square of the number of variational parameters ($P$), making their explicit construction and storage computationally infeasible when $P$ scales up, a common occurrence in modern NQS. To overcome this critical bottleneck, we propose solving the central linear algebra problems of these optimization methods by utilizing iterative Krylov subspace algorithms. Crucially, these solvers do not require the matrices to be built explicitly. Instead, they only require the efficient evaluation of matrix-vector products (MVPs), which we demonstrate to be far less computationally expensive than actually constructing the full matrices . In the context of VMC, this approach is made particularly efficient by evaluating these MVPs on-the-fly during the stochastic sampling process. Since each sampled configuration contributes an outer product to the overall matrix, operating by the matrices during sampling allows us to compute the required MVPs without ever needing to store the large $O(P^2)$ matrices, enabling us to treat massive NQS efficiently.

In this work, we will demonstrate this approach by using the implicitly restarted Lanczos (IRL) to improve the SR method. Instead of directly approximating the inverse of the ill-conditioned QGT via iterative linear solvers, we exploit the mathematical equivalence between the $\text{SR}$ update and the generalized eigenvalue problem (GEVP): $H_{\text{eff}} \, v = \lambda Sv$. We introduce the IRL as the robust, second-order solver for this GEVP. The IRL method efficiently extracts the smallest eigenvalue $\lambda_{\text{min}}$ and its corresponding eigenvector $\bm{v}_{\text{min}}$ from a small, dynamically generated Krylov subspace. This approach bypasses the ill-conditioning of $S$ and provides a step direction $v_{\text{min}}$ whose magnitude is intrinsically determined by the energy difference $\lambda_{\text{min}}$, thereby eliminating the need for external learning rate tuning ($\eta$). We demonstrate that the IRL-GEVP framework transforms NQS optimization into a numerically stable and highly efficient learning paradigm. This unprecedented stability and efficiency enables shallow NQS architectures to achieve performance previously restricted to deep or massive networks. Our results show that this approach reliably and consistently reaches chemical accuracy (and often far surpasses it), successfully bridging the long-standing gap between the theoretical promise of NQS and their practical utility in computational quantum science.

\section{Results}
\subsection{The IRL Framework for Robust Second-Order Optimization}
The variational optimization of a neural quantum state $|\psi_\theta\rangle$ aims to find parameters $\theta$ that minimize the energy $E(\theta) = \langle \psi_\theta | H | \psi_\theta \rangle / \langle \psi_\theta | \psi_\theta \rangle$. The SR update rule is derived from a first-order expansion of the imaginary-time evolution~\cite{mcardle2019variational}, leading to a linear system for the parameter update $\Delta \theta$:
\begin{equation}
    S \Delta \theta = -\eta F
    \label{eq:sr_system}
\end{equation}
where $F$ is the energy gradient vector, $S$ is the QGT, and $\eta$ is a learning rate. Solving this system via iterative methods like CG is prone to the numerical instabilities caused by the ill-conditioned matrix $S$~\cite{neuscamman2012optimizing}.

Our central methodological contribution is to reformulate this optimization task into a Hermitian eigenvalue problem. We achieve this by noting that the optimal parameter update in the SR framework can be obtained by minimizing the Rayleigh quotient of an effective Hamiltonian $\tilde{H}$ (see Methods for detailed construction). This allows us to rewrite the problem as:
\begin{equation}
    \tilde{H} v = \lambda v
\end{equation}
The lowest eigenvector $v_{\text{min}}$ of this problem yields the optimal parameter update $\Delta \theta \propto v_{\text{min}}$. Crucially, this transformation automatically determines the ideal energy descent step and direction, eliminating the manual tuning of the learning rate hyperparameter and enhancing numerical stability.

\begin{figure}[h!]
    \centering
    \includegraphics[width=0.9\linewidth]{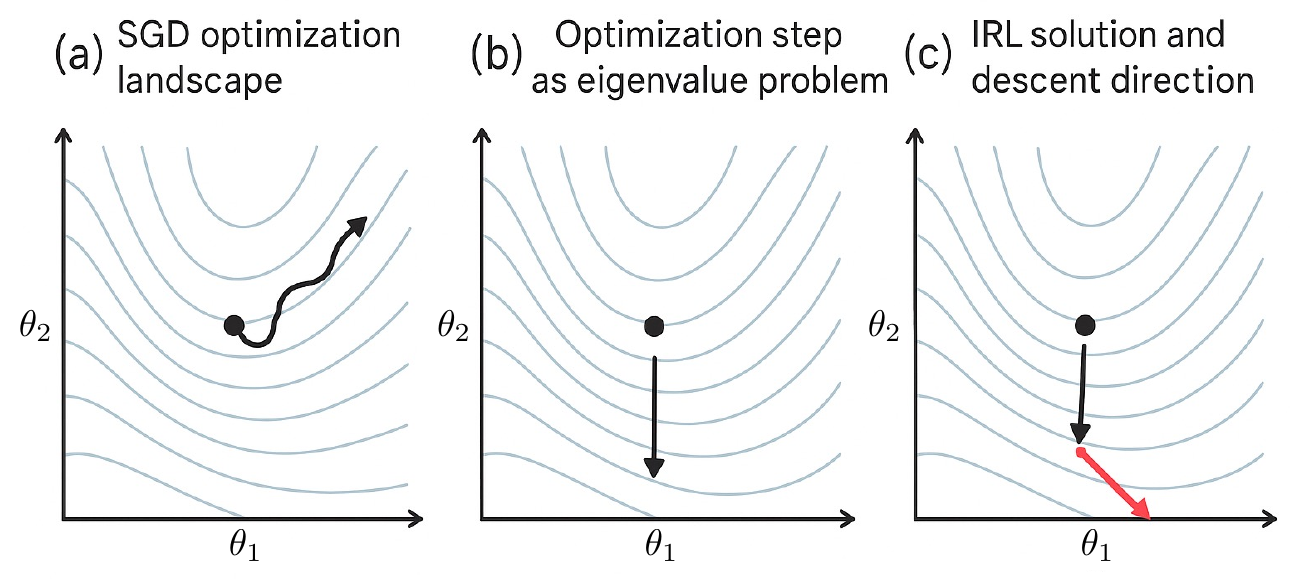}
    \caption{\textbf{Schematic of the IRL-Optimized NQS Optimization Framework.} a, The traditional first-order stochastic optimization landscape (e.g., Adam) is noisy and requires careful tuning of the learning rate ($\eta$). b, Our approach reformulates the second-order parameter update as an eigenvalue problem in the neural network's parameter space. c, The IRL method stably solves this problem, finding the optimal descent direction (red arrow) and step size automatically, enabling robust navigation of the optimization landscape.}
    \label{fig:framework}
\end{figure}

We solve this eigenvalue problem using the Algorithms~\ref{alg:irl} and \ref{alg:lanczos}. As illustrated schematically in Fig. \ref{fig:framework}, IRL iteratively builds a Krylov subspace in the parameter space, efficiently and stably converging to the extremal eigenvectors without the need for explicit matrix inversion or the severe loss of orthogonality that plagues the standard Lanczos (SL)~\cite{prelovvsek2013ground} method, particularly on high-dimensional, noisy optimization landscapes.

\subsection{Efficiency and Numerical Stability of IRL Optimization}
\begin{figure}[htbp]
    \centering
    \includegraphics[width=0.98\linewidth]{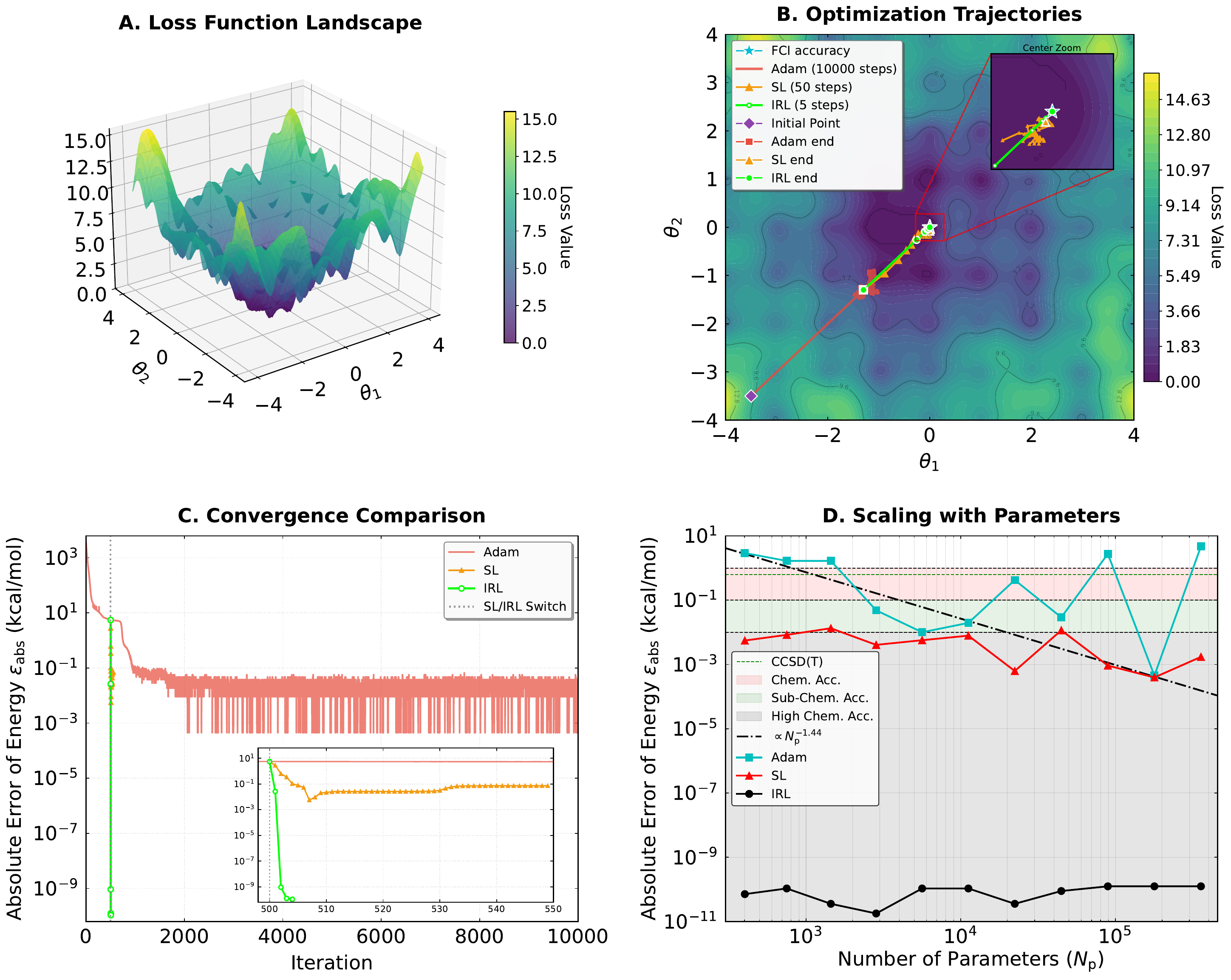}
    \caption{\textbf{Optimization landscape and trajectories for different NQS training methods, with a base neural network size of 5,622 parameters.} \textbf{(a)} Three-dimensional representation of the variational energy error surface $\varepsilon_{\text{abs}}=|E_{\text{NQS}}-E_{\text{FCI}}|$ as a function of two shifted hyperparameters $\theta_1$ and $\theta_2$, with the global minimum corresponding to FCI accuracy (the origin (0,0) is shifted to align with the global minimum for visualization clarity). \textbf{(b)} Two-dimensional projection with optimization paths for Adam, standard Lanczos (SL), and implicitly restarted Lanczos (IRL) methods. \textbf{(c)} Convergence behavior showing IRL reaching the global minimum in only 5 steps (5.1 milliseconds), compared to 10,000 Adam steps (91.2 seconds) and 50 SL steps (0.272 seconds) that fail to converge. Data are from $\text{F}_2$ molecule calculations. \textbf{(d)} Performance scaling with model size for the $\text{F}_2$ molecule. As the network dimension (e.g., number of hidden units or layers) increases, Adam's optimization behavior exhibits a clear scaling trend. In contrast, both SL and IRL demonstrate robust stability, showing minimal dependence on the model's scale. All calculations were performed with a fixed random seed of 111.}
    \label{fig:landscape}
\end{figure}

To fundamentally assess the efficiency and stability of our proposed IRL optimizer, we first visualized its optimization trajectory against conventional methods in calculations of the $\text{F}_2$ molecule. As shown in Fig. \ref{fig:landscape}a, the variational energy error landscape forms a complex, high-dimensional surface in parameter space. The trajectory of the conventional first-order Adam optimizer (Fig. \ref{fig:landscape}b, c) is tortuous and protracted, requiring over 10,000 steps to approach the global minimum and exhibiting characteristic instability. The SL method, while offering a more direct descent path, fails to converge after 50 steps due to accumulating numerical instability and loss of orthogonality. In stark contrast, the IRL optimizer demonstrates a near-ideal descent trajectory, navigating efficiently and robustly to the global minimum in merely 5 steps. This dramatic difference underscores IRL's capability to overcome the critical challenges of noise and ill-conditioning that plague other second-order iterative methods.

The practical impact is further quantified by the total wall-clock time required to achieve the highest possible accuracy for the $\text{F}_2$ molecule. The Adam optimizer required 91.2 seconds to complete 10,000 steps and failed to reach the global minimum. The SL method completed 50 steps in 0.272 seconds (a 335$\times$ speed-up over Adam) but also failed. Remarkably, the IRL optimizer achieved the global minimum in only 5.1 milliseconds, demonstrating a total runtime speed-up of approximately 17,900-fold over Adam's procedure.

This superior convergence efficiency is validated by its scaling behaviour with model size ($N_\text{p}$) (Fig. \ref{fig:landscape}d). The performance of the first-order Adam optimizer exhibits a strong dependence on $N_p$, adhering to a distinct $N_\text{p}^{-1.44}$ scaling law, indicating that achieving high accuracy with larger networks becomes progressively more difficult. Conversely, both SL and IRL demonstrate remarkable robustness, showing no significant scaling relationship with $N_\text{p}$. This independence from the model's complexity is a hallmark of a numerically stable and physically grounded optimization strategy, crucial for scaling NQS to more complex systems in quantum machine learning.

\subsection{Shallow NQS Achieve Unparalleled Accuracy and Efficiency}
We conducted a comprehensive benchmark across diverse molecules to quantify the performance of our IRL-optimized NQS. Table \ref{tab:results_summary} summarizes the results, emphasizing the absolute error ($\varepsilon_{\text{abs}}$), convergence speed (Steps), and required network complexity (Parameters, $N_\text{p}$).

The key finding is the dual advantage of extreme precision and parameter efficiency (compactness). For every system tested, our IRL-optimized shallow NQS consistently achieves extreme precision, with errors on the order of $10^{-10}$ to $10^{-12}$ $\text{kcal/mol}$. This accuracy surpasses not only all other NQS approaches but also high-level non-variational quantum chemistry methods like coupled-cluster Singles-Doubles with perturbative Triples ($\text{CCSD(T)}$)~\cite{kowalski2000method}.

More importantly for machine intelligence, this is achieved using orders of magnitude fewer parameters ($N_\text{p}$) than the large NQS optimized with Adam (e.g., $718$ parameters vs. $150,000$ for $\text{H}_2$). This confirms that the robust second-order optimization provided by IRL is highly effective at exploiting the expressive power of shallow architectures, avoiding the common machine learning strategy of needing to exponentially increase model size to compensate for poor optimization.

Furthermore, the $\text{IRL}$ optimizer converges robustly in just 3 to 5 steps for all molecules, irrespective of system size or complexity (PVCs up to $\sim 4.1\times 10^7$). This is a sharp contrast to the Adam optimizer, which necessitates 50,000 steps while yielding significantly higher errors. This systemic stability and efficiency demonstrate that IRL provides a robust and scalable learning framework for high-dimensional variational problems.

\begin{table}[h!]
\centering
\footnotesize
\caption{Performance comparison of neural quantum states (NQS) across molecular systems. For each system, we report the total orbitals ($N_\mathrm{o}$), electrons ($N_\mathrm{e}$), and number of physically valid configurations (PVCs). The shallow NQS trained with our implicitly restarted Lanczos (IRL) method achieves extreme numerical precision ($\varepsilon_{\mathrm{abs}} \sim 10^{-10}$--$10^{-12}$\,kcal/mol) in only 2--5 optimization steps, outperforming standard Lanczos (SL), Adam-optimized NQS, and CCSD(T) while using orders of magnitude fewer hyperparameters ($N_\mathrm{p}$). All calculations use the STO-3G basis set.}
\label{tab:results_summary}
\setlength{\tabcolsep}{3pt}
\begin{tabular}{@{}lccccccccccccc@{}}
\toprule
\multirow{2}{*}{Molecule} & 
\multicolumn{3}{c}{System} & 
\multicolumn{3}{c}{IRL-NQS (Ours)} & 
\multicolumn{3}{c}{SL-NQS} & 
\multicolumn{3}{c}{Adam-NQS} & 
\multirow{2}{*}{CCSD(T)} \\
\cmidrule(lr){2-4} \cmidrule(lr){5-7} \cmidrule(lr){8-10} \cmidrule(lr){11-13}
& 
$N_\mathrm{o}$ & $N_\mathrm{e}$ & PVCs & 
$N_\mathrm{p}$ & Steps & $\varepsilon_{\mathrm{abs}}$ & 
$N_\mathrm{p}$ & Steps & $\varepsilon_{\mathrm{abs}}$ & 
$N_\mathrm{p}$ & Steps & $\varepsilon_{\mathrm{abs}}$ & 
$\varepsilon_{\mathrm{abs}}$
\\
\midrule
H$_2$       & 4 & 2 & 4       & 7.2e2  & 2  & 4.6e-12 & 7.2e2  & 2    & 6.1e-5  & 1.5e5  & 5.0e4 & 8.3e-3  & 9.8e-7  \\
LiH         & 12 & 4 & 2.3e2  & 2.7e3  & 3  & 1.9e-11 & 2.7e3  & 31   & 1.7e-5  & 6.3e5  & 5.0e4 & 6.0e-2  & 3.4e-3  \\
H$_2$O      & 14 & 10 & 4.4e2  & 3.3e3  & 5  & 1.3e-10 & 3.3e3  & 100  & 8.0e-4  & 7.5e5  & 5.0e4 & 9.0e-2  & 7.6e-2  \\
H$_2$S      & 22 & 18 & 3.0e3  & 6.5e3  & 5  & 2.5e-10 & 6.5e3  & 100  & 7.9e-4  & 7.5e5  & 5.0e4 & 4.8e-1  & 4.2e-2  \\
NH$_3$      & 16 & 10 & 3.1e3  & 4.0e3  & 5  & 2.2e-11 & 4.0e3  & 100  & 9.7e-4  & 9.1e5  & 5.0e4 & 1.4e-1  & 1.4e-1  \\
CH$_4$      & 20 & 10 & 1.6e4  & 4.8e3  & 5  & 2.8e-10 & 4.8e3  & 100  & 9.2e-3  & 1.1e6  & 5.0e4 & 6.3e-2  & 1.5e-1  \\
N$_2$       & 20 & 14 & 1.4e4  & 5.6e3  & 5  & 3.3e-10 & 5.6e3  & 100  & 7.2e-5  & 9.1e5  & 5.0e4 & 3.2e-1  & 1.4      \\
LiF         & 20 & 12 & 4.4e4  & 5.6e3  & 5  & 1.3e-10 & 5.6e3  & 100  & 8.0e-4  & 1.1e6  & 5.0e4 & 9.2e-1  & 6.3e-2  \\
LiCl        & 28 & 20 & 1.0e6  & 9.6e3  & 5  & 4.8e-10 & 9.6e3  & 100  & 2.3e-3  & 1.4e6  & 5.0e4 & 1.7     & 2.5e-1  \\
Li$_2$O     & 30 & 14 & 4.1e7  & 1.1e4  & 5  & 5.8e-10 & 1.1e4  & 100  & 2.9e-3  & 1.7e6  & 5.0e4 & 1.1     & 2.7e-1  \\
\midrule
\multicolumn{4}{l}{\textbf{Parameter Efficiency}} & 
\multicolumn{3}{c}{\textbf{Same $N_\mathrm{p}$ as SL}} & 
\multicolumn{3}{c}{\textbf{Same $N_\mathrm{p}$ as IRL}} & 
\multicolumn{3}{c}{\textbf{100--200$\times$ more params}} & 
\textbf{--} \\
\multicolumn{4}{l}{\textbf{Convergence Speed}} & 
\multicolumn{3}{c}{\textbf{2--5 steps}} & 
\multicolumn{3}{c}{\textbf{2--100 steps}} & 
\multicolumn{3}{c}{\textbf{50,000 steps}} & 
\textbf{--} \\
\bottomrule
\end{tabular}
\end{table}

\subsection{Robust Optimization in the Strong Correlation Regime: Bond Breaking}
\begin{figure}[h!]
    \centering
    \includegraphics[width=0.5\linewidth]{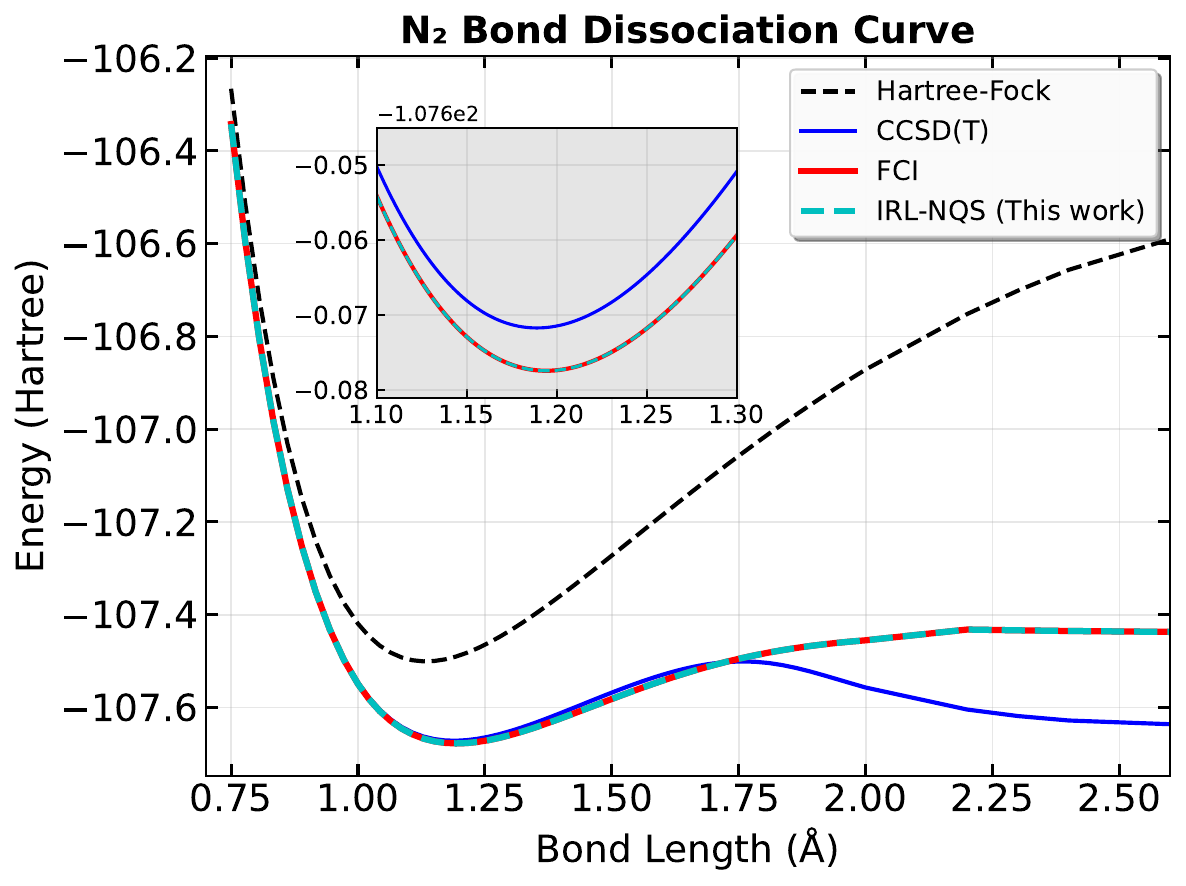}
    \caption{Bond dissociation curve of the N$_2$ molecule computed using various electronic structure methods, including Hartree–Fock, CCSD(T), FCI, and the IRL-NQS approach introduced in this work. The total energy (in Hartree) is plotted as a function of the N–N bond length (in \AA). The IRL-NQS results demonstrate close agreement with high-accuracy FCI calculations across the dissociation range.}
    \label{fig:N2}
\end{figure}

A stringent test for any high-dimensional machine learning model used in physical simulation is its stability and convergence in regions characterized by strong correlation, such as chemical bond dissociation. In optimization terms, strong correlation corresponds to a highly non-convex, multi-modal, and severely ill-conditioned loss landscape, posing an extreme challenge for standard optimization routines. We applied our IRL-NQS approach to map the bond dissociation curve of the $\text{N}_2$ molecule, a classic multi-reference benchmark. Figure \ref{fig:N2} presents the total energy as a function of the N-N bond length, comparing our shallow IRL-NQS results with Hartree-Fock~\cite{slater1951simplification}, $\text{CCSD(T)}$~\cite{kowalski2000method}, and the near-exact full configuration interaction ($\text{FCI}$)~\cite{knowles1984new} reference where available.

The comparison starkly highlights the limitations of methods reliant on single-reference approximations. $\text{Hartree-Fock}$ fails qualitatively, showing an unphysical energy rise at large separations. $\text{CCSD(T)}$, while highly accurate near the equilibrium minimum (the single-reference regime), diverges dramatically at larger bond lengths ($R>2.5R_e$) due to its inability to properly capture the complex, multi-reference electronic structure. In contrast, our IRL-NQS results demonstrate near-perfect agreement with the high-accuracy $\text{FCI}$ calculations across the entire dissociation range (see Figure \ref{fig:N2}). This robust agreement—achieved with a shallow NQS architecture—serves as definitive evidence that the IRL second-order optimizer maintains its numerical stability and convergence efficacy even when navigating the highly rugged, ill-conditioned optimization landscape of the strong correlation regime. This capability establishes the IRL-NQS framework as a numerically superior and reliable variational method for describing challenging chemical processes that require strong electron correlation capture.

\section{Discussion}
We have introduced a novel, robust second-order optimization paradigm for high-dimensional variational machine learning models by seamlessly integrating the IRL method into the SR framework. This algorithmic breakthrough directly resolves the long-standing critical issues of numerical instability and hyperparameter sensitivity that have curtailed the practical application of NQS to high-precision quantum chemistry.

The efficacy of our approach is rooted in the superior numerical properties of IRL. By reformulating the ill-conditioned natural gradient linear system $S\Delta\theta=-F$ into a stable Hermitian eigenvalue problem, IRL achieves two critical objectives: (i) The method intrinsically extracts the optimal energy descent direction and ideal step size, effectively eliminating the demanding manual tuning of the learning rate, a major bottleneck in all stochastic methods. (ii) The implicit restarting mechanism actively filters spurious eigenvectors (noise) and maintains the orthogonality of the Krylov basis vectors, preventing the convergence collapse typically seen when conventional iterative solvers encounter the highly ill-conditioned QGT. This stability is particularly crucial for training shallow network architectures, allowing them to fully realize their expressivity without relying on excessive parameter counts to compensate for poor optimization.

Our systematic benchmarking confirms that this strategy represents a fundamental algorithmic advance, moving beyond incremental improvements in convergence rate. The reliable attainment of sub-chemical accuracy ($\varepsilon_{\text{abs}} \sim 10^{-10}-10^{-12}$ kcal/mol) in just 3 to 5 steps—coupled with the ability to use shallow networks—signifies that the ultimate precision of NQS is primarily limited by the optimization method itself, not necessarily the architectural size. This dual gain in accuracy and parameter efficiency is the key to scaling NQS to larger, chemically-relevant systems where both computational cost and model complexity are primary constraints. Furthermore, the demonstrated success in the $\text{N}_2$ bond breaking curve validates IRL's generalizability to the strong correlation (multi-reference) regime, where standard single-reference methods like $\text{CCSD(T)}$ fail.

The IRL optimization framework offers a solid foundation for future variational research. Immediate applications include extending its reach to challenging chemical problems such as transition metal chemistry, excited state calculations, and molecular dynamics in non-equilibrium geometries, all of which demand the type of robust, high-fidelity convergence demonstrated here. More broadly, the success of leveraging advanced numerical linear algebra techniques, specifically the stable Krylov subspace projection, opens a new direction for optimizing complex high-dimensional machine learning models in physics and other scientific domains. Future algorithmic extensions could explore alternative Krylov subspace methods like the Davidson method~\cite{crouzeix1994davidson} or block-Lanczos methods to further exploit problem-specific structures and potentially achieve linear scaling for even larger parameter spaces.

In conclusion, our work bridges the long-standing gap between the theoretical expressivity of neural network quantum states and their practical utility, establishing IRL-NQS as a competitive, reliable, and highly efficient tool for \textit{ab initio} electronic structure theory that sets a new standard for high-precision variational optimization.

\section{Methods}
\subsection{The IRL for Wavefunction Optimization}
\label{methods:irl}

The foundation of our robust optimization strategy for NQS is the seamless integration of the IRL method within the SR framework. This innovation provides a numerically stable and efficient alternative to conventional iterative linear solvers plagued by ill-conditioning, primarily by reformulating the ill-posed parameter update problem into a well-defined Hermitian eigenvalue problem.

\subsubsection{From SR to a Stable Eigenvalue Problem}

The standard SR method is a second-order optimization technique that computes the steepest descent direction $\Delta \bm{\theta}$ in the Riemannian manifold of the NQS \cite{lin2008riemannian}. This involves solving the ill-conditioned linear system:
\begin{equation}
    \mathbf{S} \Delta \bm{\theta} = -\eta  \mathbf{F},
    \label{eq:sr_linear_system}
\end{equation}
where $\mathbf{S}$ is the QGT and $\mathbf{F}$ is the energy gradient.

We circumvent the numerical instability of inverting $\mathbf{S}$ by adopting the Linear Method (LM), which is asymptotically equivalent to SR but rooted in the Rayleigh-Ritz variational principle \cite{yserentant2013short}. The LM seeks the optimal update vector $\bm{v} = \Delta \bm{\theta}$ by minimizing the energy of the linearized wavefunction $\psi'$ in the local tangent space:
\begin{equation}
\psi'(\bm{v}) \approx \psi(\bm{\theta}) + \sum_{i=1}^{p} v_i \frac{\partial \psi(\bm{\theta})}{\partial \theta_i} = \sum_{j=0}^{p} c_j \chi_j.
\end{equation}
Here, $\bm{c} = [1, v_1, \dots, v_p]^\top$ is the coefficient vector, and $\{\chi_j\}_{j=0}^{p}$ are the basis functions $\{\psi, \partial_{\theta_1}\psi, \dots, \partial_{\theta_p}\psi\}$. The problem is to minimize the Rayleigh quotient $E[\psi']$:
\begin{equation}
E[\psi'] = \lambda = \frac{\langle \psi' | \hat{H} | \psi' \rangle}{\langle \psi' | \psi' \rangle} = \frac{\bm{c}^\dagger \tilde{\bm{H}} \bm{c}}{\bm{c}^\dagger \tilde{\bm{S}} \bm{c}}.
\label{eq:rayleigh_quotient}
\end{equation}
where $\tilde{\bm{H}}_{ij} = \langle \chi_i | \hat{H} | \chi_j \rangle$ and $\tilde{\bm{S}}_{ij} = \langle \chi_i | \chi_j \rangle$ are the full $(p+1) \times (p+1)$ Hermitian matrices.

To find the minimum $\lambda$, we apply the variational condition by taking the derivative of $E[\psi']$ with respect to $\bm{c}^\dagger$ and $\bm{c}$  and setting them to zero:
\begin{equation}
\begin{aligned}
    \frac{\partial \lambda}{\partial \bm{c}^\dagger} = 0\quad \text{and} \quad
    \frac{\partial \lambda}{\partial \bm{c}} = 0.
\end{aligned}
\end{equation}
Since the two resulting equations are equivalent under a conjugate transpose operation, they provide the same solution set for $\bm{c}$. Therefore, when deriving the GEVP, it is sufficient to impose only one of the two independent variational conditions ($\frac{\partial \lambda}{\partial \bm{c}}=0$ or $\frac{\partial \lambda}{\partial \bm{c}^\dagger}=0$) to obtain the complete GEVP form. The choice to use $\frac{\partial \lambda}{\partial \bm{c}^\dagger} = 0$ is a standard mathematical convention. 
Using the quotient rule and matrix derivatives ($N = \bm{c}^\dagger \tilde{\bm{H}} \bm{c}$, $D = \bm{c}^\dagger \tilde{\bm{S}} \bm{c}$):
\begin{equation}
\frac{1}{D^2} \left[ D \frac{\partial N}{\partial \bm{c}^\dagger} - N \frac{\partial D}{\partial \bm{c}^\dagger} \right] = 0.
\end{equation}
Substituting $\frac{\partial N}{\partial \bm{c}^\dagger} = \tilde{\bm{H}} \bm{c}$ and $\frac{\partial D}{\partial \bm{c}^\dagger} = \tilde{\bm{S}} \bm{c}$:
\begin{equation}
D (\tilde{\bm{H}} \bm{c}) - N (\tilde{\bm{S}} \bm{c}) = 0.
\end{equation}
Substituting $N/D = \lambda$ (and canceling $D$), we arrive at the GEVP:
\begin{equation}
\tilde{\bm{H}}\bm{c} = \lambda \tilde{\bm{S}}  \bm{c}.
\label{eq:full_gevp}
\end{equation}
The optimal update is given by the eigenvector $\bm{c}_{\text{min}}$ corresponding to the smallest eigenvalue $\lambda_{\text{min}}$.

To simplify the problem, we adopt the standard VMC convention that the NQS wavefunction is normalized, $\langle \psi | \psi \rangle = 1$. This ensures the current energy expectation is $E(\bm{\theta}) = \langle \psi | \hat{H} | \psi \rangle$.
To reduce the problem to the $p \times p$ parameter sub-space and ensure $\lambda_{\text{min}}$ represents the energy reduction ($\Delta E$) rather than the absolute energy ($E_{\text{new}}$), we employ the energy centering technique. We define the centered Hamiltonian operator $\hat{H}' = \hat{H} - E(\bm{\theta})\hat{I}$.

In the parameter sub-space spanned by the derivatives $\{|\partial_{\theta_i}\psi\rangle\}$, the effective Hamiltonian matrix $\mathbf{H}_{\text{eff}}$ is defined using the centered operator $\hat{H}' = \hat{H} - E(\boldsymbol{\theta})$:
\begin{equation}
\label{eqn:heff}
[\mathbf{H}_{\text{eff}}]_{ij} = \langle \partial_{\theta_i} \psi | \hat{H}' | \partial_{\theta_j} \psi \rangle = \langle \partial_{\theta_i} \psi | \hat{H} | \partial_{\theta_j} \psi \rangle - E(\boldsymbol{\theta}) \langle \partial_{\theta_i} \psi | \partial_{\theta_j} \psi \rangle.
\end{equation}
The energy expectation value is computed using the VMC technique:
$$E(\boldsymbol{\theta}) = \frac{\langle \psi | \hat{H} | \psi \rangle}{\langle \psi | \psi \rangle} = \sum_{\mathbf{x}} \frac{|\psi(\mathbf{x})|^2}{\sum_{\mathbf{x}'} |\psi(\mathbf{x}')|^2} \left[ \frac{\hat{H} \psi(\mathbf{x})}{\psi(\mathbf{x})} \right].$$
We define the local energy $\hat{E}_{\text{loc}}(\mathbf{x})$ as the ratio of the Hamiltonian acting on the wavefunction to the wavefunction itself:
\begin{equation}
\hat{E}_{\text{loc}}(\mathbf{x}) = \frac{\hat{H} \psi(\mathbf{x})}{\psi(\mathbf{x})}.
\end{equation}
The expectation value can then be expressed as the Monte Carlo average of the local energy with respect to the probability distribution $P(\mathbf{x}) \propto |\psi(\mathbf{x})|^2$: $E(\boldsymbol{\theta}) = \langle \hat{E}_{\text{loc}} \rangle_{P(\mathbf{x})}$. This local quantity is essential as it allows us to formulate the derivative of the energy, and subsequently the $\mathbf{H}_{\text{eff}}$ matrix, in a form computable through stochastic sampling.
By employing the logarithmic derivative vector $\mathbf{O}$, where $\mathbf{O}_i (\mathbf{x}) = \partial_{\theta_i} \ln \psi(\mathbf{x}; \boldsymbol{\theta})$, this relation (Eq.~\ref{eqn:heff}) can be compactly written in matrix form:
\begin{equation}
\mathbf{H}_{\text{eff}} = \langle \mathbf{O}| \hat{E}_{\text{loc}}| \mathbf{O} \rangle - E(\boldsymbol{\theta}) \mathbf{S},
\label{eq:centered_hamiltonian}
\end{equation}
where $\mathbf{S}_{ij} = \langle \mathbf{O}_i| \mathbf{O}_j \rangle$ is the QGT. This matrix $\mathbf{H}_{\text{eff}}$ is equivalent to the definition using connected correlation functions (subscript $c$), $[\mathbf{H}_{\text{eff}}]_{ij} = \langle \partial_{\theta_i} \psi | \hat{H} | \partial_{\theta_j} \psi \rangle_c$, confirming that $\mathbf{H}_{\text{eff}}$ contains only the fluctuations and correlations between the parameter space and the local energy.

This rigorous transformation leads to the final compact GEVP in parameter space, where $\bm{v}$ is the eigenvector:
\begin{equation}
    \mathbf{H}_{\text{eff}}  \bm{v} = \lambda  \mathbf{S} \bm{v}.
    \label{eq:generalized_evp}
\end{equation}
The solution to Eq.~\ref{eq:generalized_evp}, $\lambda_{\text{min}}$, strictly represents the minimum value of the Rayleigh quotient when restricted to the parameter sub-space:
\begin{equation}
\lambda = \min_{\bm{v}} \frac{\bm{v}^\dagger \mathbf{H}_{\text{eff}} \bm{v}}{\bm{v}^\dagger \mathbf{S} \bm{v}}.
\end{equation}
By substituting the definition of the energy-centered effective Hamiltonian $\bm{H}_{\text{eff}}$ into the expression for $\lambda_{\text{min}}$:
\begin{align}
\lambda_{\text{min}} &= \min_{\bm{v}} \frac{\bm{v}^\dagger \left( \langle \mathbf{O} | \hat{H} | \mathbf{O} \rangle - E(\bm{\theta}) \mathbf{S} \right) \bm{v}}{\bm{v}^\dagger \mathbf{S} \bm{v}} \\
\lambda_{\text{min}} &= \min_{\bm{v}} \left( \frac{\bm{v}^\dagger \langle \mathbf{O} | \hat{H} | \mathbf{O} \rangle \bm{v}}{\bm{v}^\dagger \mathbf{S} \bm{v}} \right) - E(\bm{\theta}).
\label{eq:lambda_min_derivation}
\end{align}
We define the term within the parentheses as $E_{\text{new}}$:
\begin{equation}
E_{\text{new}} = \min_{\bm{v}} \left( \frac{\bm{v}^\dagger \langle \mathbf{O} | \hat{H} | \mathbf{O} \rangle \bm{v}}{\bm{v}^\dagger \mathbf{S} \bm{v}} \right).
\end{equation}
$E_{\text{new}}$ represents the lowest energy attainable by the updated wavefunction $\psi'(\bm{v})$ within the linear tangent space defined by the current parameters $\bm{\theta}$.

Consequently, $\lambda_{\text{min}}$ is strictly the difference between this local minimum energy and the current energy expectation:
\begin{equation}
\lambda_{\text{min}} = E_{\text{new}} - E(\bm{\theta}).
\end{equation}

The physical interpretation of $\lambda_{\text{min}}$ is crucial for the optimization procedure. $\lambda_{\text{min}}$ represents the maximum attainable energy reduction ($\Delta E$) during the current optimization iteration, within the limits of the linear approximation. Since $\lambda_{\text{min}}$ directly provides the energy difference $\Delta E$, the method automatically determines the optimal step size. The eigenvector $\bm{v}_{\text{min}}$ defines the descent direction, and the eigenvalue $\lambda_{\text{min}}$ implicitly provides the necessary magnitude information. As the optimization progresses and the current energy $E(\bm{\theta})$ approaches the exact ground state energy $E_{\text{exact}}$, the locally minimized energy $E_{\text{new}}$ also converges towards $E_{\text{exact}}$. Consequently, the energy difference $\lambda_{\text{min}} = E_{\text{new}} - E(\bm{\theta})$ approaches zero, naturally satisfying the convergence criterion.


To leverage the efficiency and numerical stability of the IRL method, we approximate the GEVP by setting $\mathbf{S} \approx \mathbf{I}$ in the metric of the dominant descent direction, converting it into a standard Hermitian eigenvalue problem:
\begin{equation}
    \mathbf{H}_{\text{eff}} \bm{v} = \lambda  \bm{v}.
    \label{eq:standard_evp}
\end{equation}
The optimal parameter update is then determined by the eigenvector $\bm{v}_{\text{min}}$ corresponding to the smallest eigenvalue $\lambda_{\text{min}}$:
\begin{equation}
    \Delta \bm{\theta} \propto \bm{v}_{\text{min}}.
\end{equation}
The physically meaningful $\lambda_{\text{min}}$ automatically provides the optimal step size, eliminating the challenging hyperparameter tuning required by traditional SR.

\subsubsection{The Standard Lanczos Process and Implicit Restart}

We solve the SHEP (Eq. \ref{eq:standard_evp}) using the IRL algorithm. IRL builds upon the Standard Lanczos (SL) process, a projection method used to construct an orthonormal basis $\bm{V}_m = [\bm{v}_1, \dots, \bm{v}_m]$ for the $m$-dimensional Krylov subspace:
\begin{equation}
\mathcal{K}_m(\mathbf{H}_{\text{eff}}, \bm{v}_1) = \text{span}\{ \bm{v}_1, \mathbf{H}_{\text{eff}}\bm{v}_1, \dots, \mathbf{H}_{\text{eff}}^{m-1}\bm{v}_1 \}.
\end{equation}
The SL process uses a three-term recurrence relation to project $\mathbf{H}_{\text{eff}}$ onto this subspace, yielding a small, symmetric tridiagonal matrix $\bm{T}_m \in \mathbb{R}^{m \times m}$:
\begin{equation}
\bm{T}_m =
\begin{bmatrix}
\alpha_1 & \beta_1 & & & \\
\beta_1 & \alpha_2 & \beta_2 & & \\
& \beta_2 & \ddots & \ddots & \\
& & \ddots & \alpha_{m-1} & \beta_{m-1} \\
& & & \beta_{m-1} & \alpha_m
\end{bmatrix},
\end{equation}
which satisfies the projection relation $\mathbf{H}_{\text{eff}} \bm{V}_m = \bm{V}_m \bm{T}_m + \beta_m \bm{v}_{m+1} \bm{e}_m^\top$. The eigenvalues of $\bm{T}_m$ are the Ritz values, which approximate the eigenvalues of $\mathbf{H}_{\text{eff}}$.

The SL process is susceptible to numerical instability in finite-precision arithmetic, resulting in the loss of basis vector orthogonality and the appearance of spurious ghost eigenvalues. IRL addresses this by implicitly applying a polynomial filter to the Krylov subspace. This filter, defined by the unwanted Ritz values $\{\nu_1, \dots, \nu_\mu\}$ (typically those farthest from $\lambda_{\text{min}}$), effectively suppresses undesired eigencomponents:
\begin{equation}
\psi(\mathbf{H}_{\text{eff}}) = (\mathbf{H}_{\text{eff}} - \nu_1 \mathbf{I})(\mathbf{H}_{\text{eff}} - \nu_2 \mathbf{I}) \cdots (\mathbf{H}_{\text{eff}} - \nu_\mu \mathbf{I}).
\end{equation}
Instead of explicitly computing this polynomial, IRL achieves the filtering robustly by performing $\mu$ steps of the shifted QR algorithm on the small matrix $\bm{T}_m$. This procedure, guaranteed by the implicit Q theorem \cite{vandebril2005implicit, watkins1982understanding}, is mathematically equivalent to restarting the Lanczos process with an optimally filtered starting vector, ensuring numerical stability and guaranteed convergence to the extremal eigenpair $\bm{v}_{\text{min}}$.

\subsubsection{Practical Matrix-Free Implementation}

For NQS optimization, the effective Hamiltonian matrix $\mathbf{H}_{\text{eff}}$ is never explicitly constructed, as its memory requirement scales prohibitively as $\mathcal{O}(p^2)$ for $p$ parameters. Instead, the Implicitly Restarted Lanczos (IRL) algorithm operates entirely through matrix-free computations, requiring only the matrix-vector product $\mathbf{H}_{\text{eff}} \bm{v}$ at each iteration.

The $i$-th component of this product is defined theoretically as $ [\mathbf{H}_{\text{eff}} \bm{v}]_i = \sum_{j=1}^p [\mathbf{H}_{\text{eff}}]_{ij} v_j $. Using the definition of the effective Hamiltonian matrix element $ [\mathbf{H}_{\text{eff}}]_{ij} = \mathbb{E} \left[ O_i^* \left( E_{\text{loc}} - E(\bm{\theta}) \right) O_j \right] $ and the linearity of the expectation operator, the product is efficiently estimated stochastically via Monte Carlo sampling (with $N_s$ samples $\mathbf{x}_n$) without explicitly forming the matrix:
\begin{equation}
\label{eq:matrix_vec_prod}
    [\mathbf{H}_{\text{eff}} \bm{v}]_i \approx \frac{1}{N_s} \sum_{n=1}^{N_s} O_i^*(\mathbf{x}_n) \left(E_{\text{loc}}(\mathbf{x}_n) - E(\bm{\theta})\right) \left(\sum_{j=1}^p O_j(\mathbf{x}_n) v_j\right),
\end{equation}
where $E_{\text{loc}}(\mathbf{x}_n)$ is the local energy $E_{\text{loc}}(\mathbf{x}_n) = \frac{\hat{H} \psi_{\bm{\theta}}(\mathbf{x}_n)}{\psi_{\bm{\theta}}(\mathbf{x}_n)}$, and $O_i(\mathbf{x}_n) = \frac{\partial \ln \psi_{\bm{\theta}}(\mathbf{x}_n)}{\partial \theta_i}$ is the log-derivative of the wavefunction for sample $\mathbf{x}_n$. This matrix-free operation ensures that the computational cost of the IRL method scales linearly with the number of parameters $p$, $\mathcal{O}(p)$, which is essential for large-scale NQS models.

We configure the IRL algorithm to target only the lowest eigenpair ($k=1$) with a typical Krylov subspace dimension of $m=20$. A stringent convergence tolerance ($\tau=10^{-12}$) is enforced on the residual norm, ensuring the extreme chemical accuracy necessary for fundamental scientific applications. The superior efficiency of IRL is a result of its minimal required step count (typically 3--5 steps), which provides a net gain in computational speed despite a higher per-step cost compared to first-order methods.

The complete IRL procedure is summarized in Algorithm \ref{alg:irl}, which highlights the explicit construction of the Krylov subspace $\mathcal{K}_m$ through the Lanczos process. The superior overall efficiency of IRL stems from the dramatically reduced number of optimization steps required for convergence, despite a higher per-step cost.

We analyze the computational complexity in terms of the number of Monte Carlo samples ($N_s$) and the number of network parameters ($p$). For Adam/SGD, a single step requires the estimation of the energy gradient $\mathbf{F} = 2 \text{Re}(\langle \mathbf{O}^\dagger E_L \rangle_c)$, which scales as $\mathcal{O}(N_s \cdot p)$. While the core operation of the IRL (and the Lanczos process) is the Matrix-Vector product $\bm{w} = [\mathbf{H}_{\text{eff}}] \bm{v}$. Critically, we do not explicitly construct the large $\mathbf{H}_{\text{eff}}$ matrix. Instead, the product is evaluated efficiently and stochastically in $\mathcal{O}(N_s \cdot p)$ time by exploiting the sampling identity derived from the definition of connected correlations Eq.~\ref{eq:matrix_vec_prod}.

The per-step cost of IRL, $\mathbf{C_{\text{IRL}}}$, primarily involves the computation of this MVP, along with vector updates and the Rayleigh quotient calculation. In practice, we observe that the stochastic evaluation of $[\mathbf{H}_{\text{eff}} \bm{v}]$ requires slightly more computation than the stochastic evaluation of the gradient $\mathbf{F}$. This higher cost is quantified by $\mathbf{C_{\text{IRL}}} \approx \alpha \cdot \mathbf{C_{\text{Adam}}}$, where the overhead factor is typically $\alpha \in [2, 5]$.

However, the total computational cost $\text{Cost}_{\text{Total}} = \text{Steps} \times \text{Cost}_{\text{Per-step}}$ overwhelmingly favors IRL. To reach chemical accuracy, Adam often requires $\sim 50,000$ steps, while IRL converges in $\sim 5$ to $50$ steps (depending on the system complexity and Krylov subspace size $m$). This dramatic reduction in steps leads to a significant net gain in efficiency:
$$\frac{\text{Cost}_{\text{Total}}^{\text{IRL}}}{\text{Cost}_{\text{Total}}^{\text{Adam}}} \approx \frac{5 \times \mathbf{C_{\text{IRL}}}}{50000 \times \mathbf{C_{\text{Adam}}}} \approx \frac{5 \cdot \alpha}{50000} \ll 1.$$
This demonstrates that the low iteration count of the IRL method far outweighs the moderately increased cost per step, making it the superior optimizer for high-accuracy NQS computations.

\subsubsection{The Numerical Rationale for the $\mathbf{S} \approx \mathbf{I}$ Approximation}

The variational optimization of NQS relies on second-order methods, which generally aim to solve the GEVP associated with the LM or SR update: $\mathbf{H}_{\text{eff}} \bm{v} = \lambda \mathbf{S} \bm{v}$. A primary numerical challenge in this approach, particularly in large-scale NQS and strongly correlated regimes, is the severe ill-conditioning of the QGT, $\mathbf{S}$. This ill-conditioning necessitates the use of external, heuristic regularization ($\mathbf{S} \leftarrow \mathbf{S} + \epsilon \mathbf{I}$) for traditional iterative linear solvers (like $\text{CG}$) used in SR, a process that introduces its own instability and requires careful hyperparameter ($\epsilon$) tuning.

Our central methodological innovation, the $\text{IRL}$ framework, strategically employs the approximation $\mathbf{S} \approx \mathbf{I}$ to transform the optimization problem. This simplification allows us to recast the GEVP as a numerically robust SHEP: $\mathbf{H}_{\text{eff}} \bm{v} = \lambda \bm{v}$. While $\mathbf{S} \approx \mathbf{I}$ represents a theoretical departure from the exact geometric metric—particularly in strongly correlated systems where parameter redundancy is high—its rationale is rooted in a pragmatic numerical strategy that yields three critical advantages. Firstly, the transformation effectively eliminates the QGT's ill-conditioning as a computational barrier, providing the $\text{IRL}$ solver with a stable environment for extracting the extremal eigenvectors. Secondly, by solving the SHEP, the optimal parameter update direction $\mathbf{v}_{\min}$ and its magnitude (derived from the minimal eigenvalue $\lambda_{\min}$) are determined intrinsically, which entirely eliminates the need for manual tuning of the external learning rate hyperparameter—a significant source of instability in conventional methods. Finally, the $\mathbf{S} \approx \mathbf{I}$ simplification, when combined with the MVP strategy, ensures the overall complexity of the second-order update remains scalable to millions of parameters, guaranteeing computational scalability. The empirical success of the $\text{IRL}$ framework—demonstrated by its robust convergence and consistent high precision on challenging, strongly correlated systems such as the $\text{N}_2$ dissociation curve—validates this approximation as a highly effective pragmatic strategy for achieving chemical accuracy in NQS optimization.

\subsection{Wave Function Ansatz}

We employ a neural network variational ansatz formulated in the second-quantized framework to represent the many-electron wave function~\cite{liu2025absorption}. The electronic Hamiltonian in this representation is expressed as:
\begin{equation}
    \hat{H} = \sum_{pq} h_{pq} \hat{c}_p^\dagger \hat{c}_q + \frac{1}{2} \sum_{pqrs} g_{pqrs} \hat{c}_p^\dagger \hat{c}_q^\dagger \hat{c}_r \hat{c}_s,
\end{equation}
where $h_{pq}$ and $g_{pqrs}$ are one- and two-electron integrals, and $\hat{c}_p^\dagger$ ($\hat{c}_p$) are fermionic creation (annihilation) operators for a given spin-orbital $p$.

The many-body wavefunction is expanded in the Fock space as:
\begin{equation}
\ket{\psi} = \sum_{\bm{n}} \psi(\bm{n}) \ket{\bm{n}},
\end{equation}
where $\ket{\bm{n}} = \ket{n_1, n_2, \dots, n_M}$ denotes a basis state in the occupation number representation for a system with $M$ spin-orbitals, and $n_i \in \{0, 1\}$ due to the Pauli exclusion principle. The complex coefficient $\psi(\bm{n})$ is the wavefunction amplitude for configuration $\bm{n}$.

In our approach, a neural network serves as a variational function approximator for $\psi(\bm{n})$, which we denote as $\psi_{\bm{\theta}}(\bm{n})$, parameterized by $\bm{\theta}$. The network takes the occupation number vector $\bm{n}$ as input and outputs a complex number, which we decompose into its amplitude and phase:
\begin{equation}
    \psi_{\bm{\theta}}(\bm{n}) = \sqrt{p_{\bm{\theta}}(\bm{n})} \, e^{i \phi_{\bm{\theta}}(\bm{n})}.
\end{equation}
Here, $p_{\bm{\theta}}(\bm{n}) = |\psi_{\bm{\theta}}(\bm{n})|^2$ defines a probability distribution over the Hilbert space, and $\phi_{\bm{\theta}}(\bm{n})$ is the configuration-dependent phase.

\subsection{Autoregressive Modeling and Physical Constraints}

By the product rule of probability, any joint distribution $p_{\bm{\theta}}(\bm{n})$ can be factorized exactly into a product of conditional distributions:
\begin{equation}
    p_{\bm{\theta}}(\bm{n}) = \prod_{i=1}^{M} p_{\bm{\theta}}(n_i \, | \, \bm{n}_{<i}),
    \label{eq:autoregressive_factorization}
\end{equation}
where $\bm{n}_{<i} = (n_1, n_2, \dots, n_{i-1})$ denotes the occupation numbers of all spin-orbitals preceding the $i$-th orbital. This factorization allows for efficient and exact sampling of configurations $\bm{n}$ sequentially from $i=1$ to $M$, crucial for variational Monte Carlo calculations.

To ensure the physical validity of the sampled configurations, we enforce conservation of the total number of spin-up electrons ($N^\uparrow$) and spin-down electrons ($N^\downarrow$) directly within the autoregressive sampling process. For each step $i$ in the sequential sampling, the conditional probability $p_{\bm{\theta}}(n_i | \bm{n}_{<i})$ is dynamically adjusted to a restricted conditional probability $\tilde{p}_{\bm{\theta}}(n_i | \bm{n}_{<i})$ based on the current electron counts:
\begin{equation}
    \tilde{p}_{\bm{\theta}}(n_i | \bm{n}_{<i}) =
    \begin{cases}
    0, & \text{if } \mathcal{N}^\uparrow(\bm{n}_{<i}, n_i) > N^\uparrow \quad \text{or} \quad \mathcal{N}^\downarrow(\bm{n}_{<i}, n_i) > N^\downarrow, \\
    p_{\bm{\theta}}(n_i | \bm{n}_{<i}), & \text{otherwise}.
    \end{cases}
    \label{eq:constraint}
\end{equation}
Here, $\mathcal{N}^\uparrow(\bm{n}_{<i}, n_i)$ and $\mathcal{N}^\downarrow(\bm{n}_{<i}, n_i)$ are the cumulative counts of spin-up and spin-down electrons after considering the $i$-th orbital. The restricted probabilities are subsequently renormalized. This procedure guarantees that every generated sample $\bm{n}$ belongs to the subspace with fixed particle numbers $(N^\uparrow, N^\downarrow)$, drastically reducing the effective size of the Hilbert space and improving learning efficiency.

\begin{algorithm}[h!]
\caption{Implicitly Restarted Lanczos (IRL) for NQS Optimization}
\label{alg:irl}
\begin{algorithmic}[1]
\Require Initial parameters $\bm{\theta}$, desired eigenpairs $k=1$, Krylov subspace size $m=20$, tolerance $\tau=10^{-12}$
\Ensure Optimal update direction $\bm{v}_{\text{min}}$, optimal step size $\lambda_{\text{min}}$
\State Initialize random unit vector $\bm{v}_1$. Set $\bm{v}_0 \gets \bm{0}$, $\beta_0 \gets 0$ \Comment{Initialization for three-term recurrence}
\State $\mu \gets m - k$  \Comment{Number of unwanted Ritz values to shift/filter}
\State $\text{converged} \gets \text{False}$

\While{not converged}
    \State \textbf{1. Krylov Subspace Construction (Standard Lanczos):}
    \For{$j = 1$ to $m$}
        \State $\bm{w} \gets \mathbf{H}_{\text{eff}} \bm{v}_j$  \Comment{Matrix-vector product via Monte Carlo (Eq.~\eqref{eq:matrix_vec_prod})}
        \State $\alpha_j \gets \bm{v}_j^\dagger \bm{w}$ \Comment{Diagonal element $\bm{T}_{m, j, j}$}
        \State $\bm{w} \gets \bm{w} - \alpha_j \bm{v}_j - \beta_{j-1} \bm{v}_{j-1}$ \Comment{Orthogonalization (Three-term recurrence)}
        \State $\beta_j \gets \|\bm{w}\|$ \Comment{Subdiagonal element $\bm{T}_{m, j+1, j}$}
        \If {$\beta_j = 0$} \textbf{break} \EndIf \Comment{Breakdown (Invariant subspace found)}
        \State $\bm{v}_{j+1} \gets \bm{w} / \beta_j$ \Comment{Normalize to get next basis vector}
    \EndFor
    \State $\bm{V}_m = [\bm{v}_1, \dots, \bm{v}_m]$ forms the basis; $\bm{T}_m$ is the tridiagonal matrix.

    \State \textbf{2. Ritz Approximation:}
    \State Solve $\bm{T}_m \bm{U} = \bm{U} \bm{\Lambda}$ \Comment{Eigendecomposition of the small $\bm{T}_m$}
    \State $\bm{Y} \gets \bm{V}_m \bm{U}$ \Comment{Compute Ritz vectors}
    \State Select the lowest eigenpair $(\lambda_1, \bm{y}_1)$

    \State \textbf{3. Convergence Check:}
    \State $\text{res\_norm} \gets \|\mathbf{H}_{\text{eff}} \bm{y}_1 - \lambda_1 \bm{y}_1\|$ \Comment{Residual norm of the lowest Ritz pair}
    \If {$\text{res\_norm} < \tau$}
        \State $\bm{v}_{\text{min}} \gets \bm{y}_1$, $\lambda_{\text{min}} \gets \lambda_1$
        \State \textbf{break}
    \EndIf
    
    \State \textbf{4. Implicit Restart (Basis Filtering):}
    \State Select $\mu$ unwanted Ritz values $\{\nu_1, \dots, \nu_\mu\}$ as shifts.
    \State Apply $\mu$ implicit shifted QR iterations to $\bm{T}_m$.
    \State Extract the new $k$-length basis $\bm{V}_k$ and the next starting vector $\bm{v}_{k+1}$.
    \State $\bm{v}_1, \dots, \bm{v}_{k+1} \gets \text{new basis vectors}$ \Comment{Restart process with filtered basis}
\EndWhile
\State \Return $\bm{v}_{\text{min}}$, $\lambda_{\text{min}}$
\end{algorithmic}
\end{algorithm}

\begin{algorithm}[h!]
\caption{Standard Lanczos (SL) Process for Tridiagonal Reduction}
\label{alg:lanczos}
\begin{algorithmic}[1]
\Require Hermitian matrix $\bm{H}_{\text{eff}}$ (implicit), initial unit vector $\bm{v}_1$, subspace size $m$
\Ensure Orthonormal basis $\bm{V}_m = [\bm{v}_1, \dots, \bm{v}_m]$, tridiagonal matrix $\bm{T}_m$
\State $\beta_0 \gets 0$, $\bm{v}_0 \gets \bm{0}$ \Comment{Initialize auxiliary variables}
\For{$j = 1$ to $m$}
    \State $\bm{w} \gets \bm{H}_{\text{eff}} \bm{v}_j$ \Comment{Matrix-vector product via Eq.~\eqref{eq:matrix_vec_prod}}
    \State $\alpha_j \gets \bm{v}_j^\dagger \bm{w}$ \Comment{Rayleigh quotient: $\alpha_j = \langle \bm{v}_j, \bm{H}_{\text{eff}} \bm{v}_j \rangle$}
    \State $\bm{w} \gets \bm{w} - \alpha_j \bm{v}_j$ \Comment{Orthogonalize against current basis vector}
    \If{$j > 1$}
        \State $\bm{w} \gets \bm{w} - \beta_{j-1} \bm{v}_{j-1}$ \Comment{Orthogonalize against previous basis vector}
    \EndIf
    \State $\beta_j \gets \|\bm{w}\|$ \Comment{Compute subdiagonal element}
    \If{$\beta_j = 0$} \Comment{Invariant subspace found (lucky breakdown)}
        \State \textbf{break}
    \EndIf
    \State $\bm{v}_{j+1} \gets \bm{w} / \beta_j$ \Comment{Normalize to obtain next basis vector}
\EndFor
\State \Return $\bm{V}_m$, $\bm{T}_m$ with elements $\{\alpha_1, \dots, \alpha_m\}$ and $\{\beta_1, \dots, \beta_{m-1}\}$
\end{algorithmic}
\end{algorithm}

\begin{acknowledgement}

This work is supported by the National Natural
Science Foundation of China (Nos. 22273075 and 22361142829), the Zhejiang Provincial Natural Science
Foundation (No. XHD24B0301).

\end{acknowledgement}


\bibliography{ref}

\end{document}